\documentstyle[fleqn]{article}

\begin{document}
\title{  Thermodynamic properties of the exactly solvable transverse
             Ising model on  decorated planar lattices}
\author{Jozef Stre\v{c}ka and  Michal Ja\v{s}\v{c}ur\\
\normalsize Department of Theoretical Physics and Astrophysics,
        Institute of Physics
\\
\normalsize P. J. \v{S}af\'{a}rik University,
Moyzesova 16, 041  54 Ko\v{s}ice, Slovak Republic\\
\normalsize E-mail:jozkos@pobox.sk, jascur@kosice.upjs.sk}
\date{}
\maketitle
%**************************************************************
\begin{abstract}
The generalized mapping transformation technique
is used to obtain the exact solution for the transverse Ising model on
decorated planar lattices. Within this scheme,
the basic thermodynamic quantities
are calculated for different planar lattices
with arbitrary spins of decorating atoms. The particular
attention has been paid to the investigation of transverse-field
effects on magnetic properties of the system under
investigation. The most interesting numerical results for the
phase diagrams, compensation temperatures and several
thermodynamic quantities are discussed in detail for the
ferrimagnetic version of the model.
\end{abstract}
%**************************************************************
\section{Introduction}
%*************************************************************
For many years magnetic properties of decorated
Ising spin systems consisting of spin-$1/2$ and spin-$S$ ($S \geq 1/2$)
atoms, have been very intensively studied both theoretically and experimentally.
In particular, the behaviour of various decorated models has been explored
by variety of mathematical techniques, with some exact results \cite{1},
even in the presence of the single-ion anisotropy \cite{2}.
The strong interest in these models arises partly on account
of the rich critical phenomena they display and partly
due to the fact that they represent more complicated but
simultaneously exactly solvable systems than the simple, undecorated
ones.
Moreover, the decorated planar models belong to the simplest
exactly solvable theoretical models of ferrimagnetism. In this respect,
they may exhibit under certain conditions the compensation phenomenon,
i. e. the compensation point at which the resultant magnetization
vanishes bellow the critical temperature.
All these properties of decorated systems make them very interesting also from
the experimental point of view, first of all in connection with many
possible technological application of ferrimagnets to practice.
On the other hand, much work is currently being done in the
area of molecular-based
magnetic materials and so molecular magnetism
has become an important topic of scientific interest.
In fact, a rapid progress in molecular engineering brought the
possibility to control and to design the magnetic structure and
properties of molecular systems. Thus, a number of bimetallic
network assemblies have been synthesized which perfectly fulfil the claim
of sufficiently small interplane coupling. Nevertheless,
the molecular magnets which would possess the decorated network
structure have been prepared only a decade
ago. From this family, the most frequently prepared compounds
are the bimetallic assemblies having the decorated honeycomb
sheet structure \cite{4} or the decorated square sheet structure
\cite{5} (see Fig. 1). Obviously, the experimental discovery of this
wide class of compounds has stimulated the renewed interest in
studying the mixed spin-$1/2$ and spin-$S$ ($S \geq 1/2$) decorated
planar models theoretically.

Owing to these facts, the present article will be devoted to the
investigation one of the simplest mixed-spin quantum model.
Namely, we will study the decorated model in the presence of a transverse field
that can potentially be useful for understanding some of the above
 mentioned experimental systems.
The transverse Ising model
has been originally introduced by de Gennes as a pseudo-spin model of
hydrogen bonded ferroelectrics \cite{6}, however, during the last
decade it has found a wide application in the description of
diverse physical systems, for instance, cooperative
Jahn-Teller systems \cite{7}, strongly anisotropic magnetic materials
in the transverse field \cite{8}, or many other systems \cite{9}.
Although the transverse Ising model is one of the simplest quantum models,
the complete exact solution has been obtained in the one-dimensional
case  only \cite{10}. For two-dimensional systems there are exactly known
only the initial transverse susceptibilities on regular
lattices \cite{11}. On the other hand, over the last few years
a simple straightforward method has been developed for obtaining
exact results to the transverse Ising model, by assuming that it is
composed of quantal and 'classical' (Ising-type) spins \cite{12, 13}.
The  method is based on the generalized
mapping transformation (the decoration-iteration or star-triangle one)
introduced into the system with the transverse field affecting only
one kind of spins (quantal).

The primary purpose of this work is to provide
further extension of the exact solution from the reference \cite{12}, by the use
of the same approach. Adopting the basic ideas of the
transformation techniques, we examine the influence of
applied transverse field on the thermodynamic properties of decorated
lattices with different planar topology and arbitrary spin values of
decorating atoms.

The outline of this paper is as follows. In Section 2, the
fundamental framework of the transformation method as applied to
the present model, are briefly reviewed. This is followed by a
presentation of the numerical results for the
spin-$\mu$ ($\mu = 1/2$) and spin-$S_B$ ($S_B \geq 1/2$) decorated transverse Ising
ferrimagnet on several planar lattices, in Section 3.
Finally, some concluding remarks are given in Section 4.

%**************************************************************
\section{Formulation}
%*************************************************************
In this work we will study the transverse Ising model on decorated
planar lattices. A typical example of the system under
investigation is depicted for the case of the square lattice in
Fig. 1.  As one  can see, the system consists of two
interpenetrating sublattices
 and we will further assume that the sites of the
original lattice that constitute the sublattice $A$ are occupied
by atoms with the fixed spin $\mu = 1/2$. The second sublattice $B$
is occupied by decorating atoms with an arbitrary
spin value $S_B$. Then the system is described by the Hamiltonian
\begin{equation}
\hat {\cal H}_d = - \frac12 \sum_{i, k} J \hat S_i^z \hat \mu_k^z  -
 H_A \sum_{k=1}^{N} \hat \mu_k^z - H_B \sum_{i=1}^{Nq/2} \hat S_i^z -
\Omega \sum_{i=1}^{Nq/2} \hat S_i^x,
\label{eq1}
\end{equation}
where $\hat \mu_k^z, \hat S_i^z$ and $\hat S_i^x$ represent the relevant
components of standard spin operators and $J$ is the exchange
integral that specifies the exchange interaction between
nearest-neighboring atoms. The last three terms
describe the interaction of $A$
atoms with an external longitudinal magnetic field $H_A$, as well
as the interaction of $B$ atoms with an external longitudinal ($H_B$)
and transverse ($\Omega$) magnetic field, respectively.
Furthermore, $N$ and $q$ denote the total number of atoms and
the coordination number  of the original lattice, respectively.
As we have already mentioned, the transverse Ising model in which
all the atoms interact with the transverse field is not exactly
solvable for two- and three-dimensional lattices.
This fact is closely related to the
mathematical complexities appearing due to the
noncomutability of the spin operators in the relevant Hamiltonian.
However, in our recent work \cite{12} we have
succeeded to solve exactly the simplified version of this model
(described by the Hamiltonian (\ref{eq1}))
in which only the decorating ($B$) atoms interact with the
transverse field. In this section, we at first briefly repeat the main
formulae derived in the Ref. \cite{12} and then we further extend the
formalism for the calculation of several interesting
thermodynamic quantities of this model.

Following the same steps as in Ref.
\cite{12}, one obtains a simple relation between the partition
function of the decorated transverse Ising model $({\cal Z}_d)$ and that of the
spin-1/2 Ising model on the original lattice $({\cal Z}_0)$. Namely,
\begin{equation}
{\cal Z}_d(\beta, J, \Omega, H_A, H_B) = A(\beta, J, \Omega, H_B)^{Nq/2}
                     {\cal Z}_0(\beta, R, H),
\label{eq2}
\end{equation}
where $\beta = 1/k_B T$, $k_B$ is Boltzmann constant and $T$
is the absolute temperature. The parameters $A, R$ and $H$
represent  so-called decoration-iteration parameters
that specify the corresponding original (undecorated) system
as well as decorating system. All these parameters can be obtained
from the decoration-iteration transformation (see Ref. \cite{12}) and they
are, respectively, given by
\begin{eqnarray}
 A &=& (V_1 V_2 V_3^2)^{1/4}, \nonumber \\
 \beta R &=& \ln \frac{V_1 V_2}{V_3^2}, \nonumber \\
 \beta H &=& \beta H_A + \frac{q}{2} \ln \frac{V_1}{V_2},
\label{eq3}
\end{eqnarray}
where we have defined the functions $V_1, V_2$ and $V_3$ as
follows:
\begin{eqnarray}
  V_1 &=& \sum_{n = -S_B}^{S_B} \cosh \Bigl(
             \beta n \sqrt{(J+H_B)^2 + \Omega^2}
                                      \Bigr), \nonumber \\
  V_2 &=& \sum_{n = -S_B}^{S_B} \cosh \Bigl(
             \beta n \sqrt{(J-H_B)^2 + \Omega^2}
                                      \Bigr), \nonumber \\
  V_3 &=& \sum_{n = -S_B}^{S_B} \cosh \Bigl(
             \beta n \sqrt{H_B^2 + \Omega^2}
                                      \Bigr).
\label{eq4}
\end{eqnarray}
It is worth noticing that the foregoing equations   express
the exact mapping relationship between the decorated model
under investigation and the spin-$1/2$ model on the corresponding
original lattice. One should also notice that this transformation
is valid for arbitrary values of the decorating spin $S_B$
in the presence of the external longitudinal fields.
Thus, the relevant thermodynamic quantities can be
calculated even for the nonzero longitudinal fields.
Unfortunately, after deriving the final equations,
 we have  to set $H_A = H_B = 0$ since the original planar
Ising models are exactly solvable for $H=0$ only.

In order to investigate the magnetic properties of the system,
we employ the well-known relations from the statistical mechanics
and thermodynamics. Indeed, after a simple calculation  one
derives the following equations for the Gibbs free energy, internal
energy, enthalpy, entropy and specific heat:
\begin{equation}
{\cal G}_d = {\cal G}_0 - \frac{Nqk_BT}{2} \ln A,
\label{eq5}
\end{equation}
\begin{equation}
{\cal U}_d = \Bigl( \frac{2 {\cal U}_0}{R} - \frac{Nq}{4} \Bigr)
\frac{J^2}{\sqrt{J^2 + \Omega^2}} K_0,
\label{eq6}
\end{equation}
\begin{equation}
H_d = \frac{2 {\cal U}_0}{R}
               \Bigl( \sqrt{J^2 + \Omega^2}K_0 - \Omega K_1 \Bigr) -
\frac{Nq}{4} \Bigl( \sqrt{J^2 + \Omega^2}K_0 + \Omega K_1   \Bigr),
\label{eq7}
\end{equation}
\begin{eqnarray}
{\cal S}_d = {\cal S}_0 \! \! &+& \! \! \frac{Nqk_B}{2} \ln A
- \frac{Nq}{4T} \Bigl( \sqrt{J^2 + \Omega^2} K_0 + \Omega K_1 \Bigr), \nonumber \\
 \qquad {\cal S}_0 \! \! &=& \! \! k_B \ln {\cal Z}_0 + \frac{2 {\cal U}_0}{R T}
                      \Bigl( \sqrt{J^2 + \Omega^2} K_0 - \Omega K_1
\Bigr),
\label{eq8}
\end{eqnarray}
and
\begin{equation}
{\cal C}_{\Omega} = {\cal C}_0
+ \frac{Nq}{4 k_B T^2}
\biggl [ (J^2 + \Omega^2)(K_2 - K_0^2)
         + \Omega^2 (K_3 - K_1^2)
\biggr ],
\label{eq9}
\end{equation}
where ${\cal G}_d$,  ${\cal U}_d$,  $H_d$, ${\cal S}_d$ and
${\cal C}_{\Omega}$ denote, respectively,  the Gibbs free energy, internal energy,
enthalpy, entropy and specific heat of the decorated lattice.
Similarly, ${\cal G}_0$, ${\cal U}_0$, ${\cal S}_0$ and
${\cal C}_0$ (${\cal C}_0 = T (\partial {\cal S}_0 / \partial
T)_{\Omega}$) represent the relevant quantities  of the  original lattice
that are well-known for two-dimensional systems \cite{14}.
Finally,  the coefficients $K_0-K_3$ depend on the
temperature and transverse field and they are are given by
\begin{equation}
K_0 = F_1(J,\Omega), \;
K_1 = F_1(0,\Omega), \;
K_2 = F_2(J,\Omega), \;
K_3 = F_2(0,\Omega),
\label{eq10}
\end{equation}
where
\begin{eqnarray}
F_1(x,y) = \frac{\displaystyle \sum_{n = -S_B}^{S_B}
n \sinh (\beta n \sqrt{x^2 + y^2})
           }
{\displaystyle \sum_{n = -S_B}^{S_B} \cosh (\beta n \sqrt{x^2 +
y^2})
           },
\label{eq11}
\end{eqnarray}
\begin{eqnarray}
F_2(x,y) = \frac{\displaystyle \sum_{n = -S_B}^{S_B}
n^2 \cosh (\beta n \sqrt{x^2 + y^2})
           }
    {\displaystyle \sum_{n = -S_B}^{S_B} \cosh (\beta n \sqrt{x^2 +
y^2})
           }.
\label{eq12}
\end{eqnarray}

Next, we turn to the calculation of the magnetization.
The spontaneous longitudinal magnetization, as well as transverse
one can be directly obtained by the differentiation of the
Gibbs free energy (\ref{eq5}) with
respect to the relevant longitudinal and transverse magnetic fields.
Consequently, the spontaneous longitudinal magnetization per spin
can be written for both sublattices in the following compact form:
\begin{eqnarray}
m_A^z &=& m_0, \nonumber \\
m_B^z &=& m_0 \frac{2J}{\sqrt{J^2+ \Omega^2}} K_0,
\label{eq13}
\end{eqnarray}
where $m_A^z$  $(m_B^z)$ means the spontaneous longitudinal magnetization
of sublattice $A$ $(B)$ and $m_0$ stands for the spontaneous magnetization of
the original lattice ($m_0$ depends only on the exchange
interaction $R$ and temperature).  Similarly, the same procedure
leads to the simple relation for the reduced transverse magnetization
of sublattice $B$. Namely,
\begin{equation}
m_B^x = \frac12 \biggl (
 \frac{\Omega}{\sqrt{J^2+ \Omega^2}} K_0 + K_1
                \biggr ) +
        2 \varepsilon_0  \biggl (
 \frac{\Omega}{\sqrt{J^2+ \Omega^2}} K_0 - K_1
                \biggr ).
\label{eq14}
\end{equation}
Here, $\varepsilon_0$ denotes the two-spin correlation function
between nearest-neighboring spins of the original lattice
(depending again only on the exchange parameter $R$ and
temperature).
To complete the analysis of the  system under investigation,
one has to investigate the phase boundaries, as well as
a possibility of  compensation phenomena in the case of
the ferrimagnetic ordering ($J<0$).
The structure of equations for the spontaneous sublattice magnetization
implies that the phase transition temperature $T_c$
(or $\beta_c= 1/(k_B T_c)$) can be directly found from the
transformation formula after setting $H_A = H_B = H = 0$ and
substituting the inverse critical temperature $\beta_c R$ of the relevant original lattice into (\ref{eq3}).
In this way one obtains the relation
\begin{equation}
\beta_c R =
            2 \displaystyle \ln \frac{\displaystyle \sum_{n = -S_B}^{S_B} \cosh \Bigl(
             \beta_c n \sqrt{J^2 + \Omega^2}            \Bigr) }
{\displaystyle \sum_{n = -S_B}^{S_B} \cosh \Bigl(
             \beta_c n \Omega \Bigr) },
\label{eq15}
\end{equation}
which is valid for any planar lattice with the decorating spin $S_B$
and for the special case of $S_B = 1/2$
naturally recovers Eq.(13) in the Ref. \cite{12}.

Moreover, in the case of  ferrimagnetic system
the compensation temperature $T_k$
(or $\beta_k=1/(k_B T_k) $) can be found from the condition
that the total magnetization $M$ of the system vanishes bellow the
critical temperature. Since in our case we have $M= N( m_A +
qm_B/2)$, then from the condition $M=0$ one easily finds the
following relation for the compensation temperature
\begin{equation}
\frac{q|J|}{\sqrt{J^2 + \Omega^2}} K_0 (\beta_k) = 1,
\label{eq16}
\end{equation}
where the inverse compensation temperature $\beta_k$ is also
included in the coefficient $K_0$.

Finally, we determine the transverse susceptibility of
the decorated transverse Ising system. For this aim, we
differentiate  the formula (\ref{eq14}) for the transverse
magnetization with respect to the transverse field and
after a straightforward but a little bit tedious algebra,
we can write the transverse
susceptibility $\chi_T$ in the form
\begin{eqnarray}
\chi_T &=&   \frac12 \biggl \{
     \frac{J^2}{(J^2 + \Omega^2)^{3/2}} K_0 +
     \frac{\beta \Omega^2}{J^2 + \Omega^2} (K_2 - K_0^2)
      + \beta(K_3 - K_1^2)
                \biggr \}    \nonumber \\
 &+&          2 \varepsilon_0  \biggl \{
     \frac{J^2}{(J^2 + \Omega^2)^{3/2}} K_0 +
     \frac{\beta \Omega^2}{J^2 + \Omega^2} (K_2 - K_0^2)
      - \beta(K_3 - K_1^2)
                \biggr \}   \nonumber \\
 &+&      4 \frac{\partial \varepsilon_0}{\partial R}
                \biggl \{
     \frac{\Omega}{\sqrt{J^2 + \Omega^2}} K_0 - K_1
                \biggr \}^2.
\label{eq17}
\end{eqnarray}
In above, $\varepsilon_0$ means the two-spin correlation function between
nearest-neighboring atoms of the original lattice

%**************************************************************
\section{Numerical results}
%*************************************************************

In this section we will illustrate the effect of the transverse
field, as well as the influence of the decorating spin $S_B$
on  magnetic properties of the system under investigation.
Moreover, the role of the lattice topology will be also examined.
Although we will restrict our numerical calculation
to the ferimagnetic case ($J<0$) only, the detailed investigation
reveals that all  dependences (excepting those for the
longitudinal magnetization)  remain unchanged also for the
ferromagnetic version of the model. This observation follows from
the fact that the relevant equations are independent
under transformation $J\to -J$.

We start our discussion with the analysis of the critical and
compensation temperatures.
At first, the variations of the critical (dashed lines)
and compensation temperatures (solid lines) with the transverse
field are shown in Fig. 2. Here, we have selected the system with
$q=4$ (i.e. the decorated square lattice), taking different spin values
of decorating atoms ($S_B$). On the other hand, in Fig. 3 we have
illustrated the influence of the lattice
topology (of different $q$) on the critical and compensation
temperature for the system with the fixed decorating spin ($S_B =1$).
In both figures, the ordered ferrimagnetic phase is stable bellow
dashed lines, and the disordered paramagnetic one becomes stable above
the relevant boundary. A closer mathematical analysis reveals
that the phase transition between these two phases is of the second
order and belongs to the same universality class as that of the
usual spin-1/2 Ising model.
As one can  see, the qualitative features of
the results do not significantly depend neither on the lattice topology nor the
spin value of atoms of sublattice $B$. In fact,
the critical temperature monotonically decreases with
increasing in the transverse field, but only in the limit of the infinity
strong transverse field tends to zero (due to the fact that
the transverse field affects only  one sublattice).
One also observes here that the value of transition point
increases with the coordination number of the original lattice, as well as
with the spin value of decorating atoms.
Contrary to this behavior,  the
compensation temperature seems to be independent of the
transverse field strength (with the accuracy of twelve orders),
although it changes both with the coordination number $q$ and the
spin value $S_B$.
In general, on basis of our numerical calculation one can state
that for arbitrary but fixed $q$ and $S_B$ the compensation
points appear only for $\Omega_k/|J| = \sqrt{q^2S_B^2 - 1}$. It
is easy to find
that this characteristic value of $\Omega$ can be obtained
from the Eq. (\ref{eq16}) by taking the limit $T_k\to 0$ (or
$\beta_k \to \infty$). Physically, the independence of the compensation
temperature on the transverse field comes from the
fact that the compensation effects appear at relatively strong
transverse fields where the relevant  transition temperature is
rather low and this fact significantly influences the
behavior of sublattice magnetization.

Now, we turn to the discussion of the internal
energy and enthalpy.
Owing to the fact, that all decorated lattices behave similarly,
we will further present numerical results for one representative lattice, namely, the decorated square
lattice.
In Fig. 4 we have depicted
the thermal variations of the internal energy and enthalpy
for the spin case $S_B = 1/2$ ($N_t$ denotes a total number of atoms).
From these dependences one finds that both quantities tend monotonically
to zero with increasing the temperature. Nevertheless, if we compare
both dependences, we can conclude that the enthalpy is more sensitive (changes the shape of the
curve more rapidly) to the transverse field than the
internal energy.
It is also clear that the thermal dependences
of the internal energy (as well enthalpy) exhibit a typical weak
energy-type singularity
behaviour irrespective of the strength of transverse field.

Further,  we have also
examined the temperature and  transverse field dependences of the
spontaneous longitudinal magnetization.
In Figs. 5 and 6 we report some typical results
for the total spontaneous longitudinal magnetization,
when the value of the transverse field is changed
(we choose the magnetization curves at temperatures
$k_BT/J = 0.1$ and $k_BT/J = 0.2$ and different spin values of
decorating atoms from $1/2$ until $2$). It follows from these
dependences that the total magnetization may exhibit two possible
shapes of the magnetization curves. Namely,  the magnetization curve with one
compensation point (at lower temperatures) and the downward magnetization
curvature without any compensation points (at higher temperatures).
Both types of the magnetization curves are closely related to the
fact that the transverse field does not directly act on the atoms of
original lattice. Hence, the spontaneous magnetization of
sublattice $A$ varies very smoothly with the transverse field,
whereas the spontaneous magnetization of sublattice $B$
is rapidly destroyed with the transverse field increasing.
On the other hand, the temperature dependences of the spontaneous
magnetization of both sublattices are  standard,
regardless of the transverse field strength. The only
exceptional case arises as the transverse field reaches the
value $\Omega_c$ at which the considered system behaves similarly
as an ordinary antiferromagnet.

In contrast to the standard temperature variations of the
longitudinal magnetization, the transverse magnetization may
display very interesting and unexpected thermal behaviour. To
illustrate the case, we have depicted in Figs. 7 and 8 some typical
temperature dependences of the transverse magnetization.
As shown in Fig. 7 for the spin case $S_B = 2$, the
transverse magnetization at relatively small transverse field
($\Omega / J = 1.5$) firstly gradually decreases to its local
minimum value and then nearby the transition point increases
in the narrow temperature region. However, the transverse magnetization
by the stronger transverse fields ($\Omega / J = 2.0$ and $3.0$)
remains almost constant and again in the vicinity of the Curie point
gradually increases, too. Far beyond the transition temperature, the transverse
magnetization monotonically decreases with increasing in temperature,
regardless of the transverse field
strength. The possible explanation of
the temperature-induced increase of the transverse magnetization can be
related to the spin release from the spontaneous magnetization
direction and spin-ordering towards the transverse field
direction. In fact, the stronger the transverse field, the
smaller the transition temperature and therefore, by sufficiently
strong transverse field the transverse magnetization
increases from its initial value, since the thermal
reshuffling is relatively small in this temperature region.
On the other hand, when the transverse field is smaller,
the relevant spin reorientation takes place at higher
temperatures,  thus the transverse magnetization
firstly decreases to its local minimum due to the
strong thermal fluctuations in this temperature region. Apparently,
also the increase in the transverse magnetization which is connected
with the spin reorientation is then smaller, since it is overlapped
with the stronger thermal fluctuation.
Before proceeding further, we have depicted in Fig. 8
the transverse magnetization against the temperature that
illustrate the influence of the different
values of spin variable $S_B$.

Furthermore, let us now look more closely at the thermal
variations of the specific heat For this  purpose, we have studied
the specific heat of the simplest spin case $S_B = 1/2$.
As one can see from Fig. 9, the logarithmical
singularity in the specific heat dependence indicates the second order
phase transition towards the paramagnetic state.
Moreover, it can be also
clearly seen that the broadening of the maximum
in the paramagnetic region of the specific heat arises
due to the transverse field effect. It turns out that the observed
maximum may be thought as a Schottky-type maximum, which has its
origin in the thermal excitation of the paramagnetic spins
inserted into the transverse field.
Next, in order to illustrate the effect of increasing spin $S_B$
we have shown in Fig. 10 the thermal dependences of specific heat
for several  values of the spin variable $S_B$. As one can ascertain,
the lower the spin value $S_B$ of decorating atoms, the
stronger the influence of the transverse field on the excitation of
these spins in the paramagnetic region. To complete our analysis
of the specific heat, we have plotted in Fig. 11 the
transverse-field dependences of the specific-heat
for $S_B = 1/2$.
As one can  expect, the depicted
behaviour strongly depends on the temperature and above
the critical point,  the dependence changes suddenly due to the
paramagnetic character of the system.

Moreover, in Fig. 12 we display the transverse susceptibility against
the transverse field for the same spin case as in Fig. 11.
Here one observes, that for sufficiently high temperature (see
the dashed line), the transverse susceptibility
falls down monotonically with the increasing  transverse field.
Contrary to this behavior, for the lower temperature (see the solid
line), the  transverse susceptibility exhibits the standard singularity
due to the second order phase transition from the ferrimagnetic state to the
paramagnetic one. For completeness, in the insert of Fig. 12, we compare
the variations of the transverse susceptibilities for different
spin values of decorating atoms. Apparently, there is no
essential difference in the behaviour of the
transverse susceptibilities, although it is shifted to higher
temperatures, as the spin value of the decorating atom is
increased. Finally, in Fig. 13 we have depicted the
temperature dependence of transverse susceptibility for
some selected values of the transverse field. Here one can see,
that for the smaller
transverse fields, for instance ($\Omega / J = 0.5$),
the transverse susceptibility
decreases with increasing the temperature, then diverges at $T_c$
and repeatedly decreases. However, in the case of the
stronger transverse fields ($\Omega / J = 1.0$ and $1.5$),
the transverse susceptibility remains at its initial
value, exhibits a divergence at phase transition and afterwards
whether gradually decreases (the case $\Omega / J = 1.0$)
or exhibits a broad maximum (the case $\Omega / J = 1.5$).
The existence of this broad maximum
in the paramagnetic region arises evidently on account of the thermal
excitation of the paramagnetic spins inserted into the transverse
field and therefore, can be observed for relatively strong
transverse fields only.

%**************************************************************
\section{Conclusion}
%*************************************************************

In this work we have presented the exact results (the phase diagrams,
compensation temperatures, spontaneous longitudinal magnetization,
transverse magnetization, internal and free energy, enthalpy, entropy,
specific heat and transverse susceptibility) for the ferrimagnetic
transverse Ising model on decorated planar lattices.
We have illustrated that the magnetic properties
of the  system under investigation exhibit the characteristic
behaviour depending on the strength of the applied transverse
field and the spin of the decorating atoms.
In particular, we have found that the considered
ferrimagnetic system does not exhibit more than one compensation
temperature that is surprisingly
completely  independent of the transverse field, though it
depends on the coordination number of the original lattice and
also on the spin of the decorating atoms.
As far as we know, such a finding has not been reported in the
literature before. Perhaps, the most interesting result to
emerge here is the temperature-induced increase of the transverse
magnetization in the vicinity of the transition temperature.
We have found a strong evidence that this increase arises due to
the spin release from the spontaneous magnetization direction,
since in the vicinity of the transition temperature spins tending to align
into the transverse field direction.

Finally, we would like to emphasize that the presented method can
by applied to more complex and realistic models, for example,
the transverse Ising models with a crystal field anisotropy,
or those with next-nearest-neighbor and multispin interactions.
Further generalizations are possible by increasing the spin of
atoms on the sublattice $A$ or introducing more realistic Heisenberg
interactions. In addition to the above mentioned
generalizations, one can also  obtain very accurate results for
this model on three-dimensional lattice. This can be done
by combining the present method with other accurate methods,
such as series expansion technique, Monte Carlo simulations  or
renormalization group methods.

Acknowledgement: This work has been supported by the Ministry of
Education of Slovak Republic under VEGA grant No. 1/9034/02.

\newpage

{\bf Figure captions}
\begin{itemize}
\item [Fig.1]
Part of the decorated square lattice. The black circles denote the
spin-1/2 atoms of sublattice $A$ (referred to as the atoms of original lattice)
and gray circles represent the spin-$S_B$ atoms of sublattice $B$ (decorating
atoms).
\item [Fig.2]
Phase boundaries (dashed lines) and the compensation
temperatures (solid lines) in the $\Omega$ - $T$ plane for the
decorated square lattice ($q=4$) and different spin values of the
decorating atoms.
\item [Fig.3]
Critical (dashed lines) and compensation (solid lines) temperatures as a
function of the transverse field for various decorated planar
lattices with the fixed decorating spin $S_B = 1$.
\item [Fig.4]
Thermal variations of the internal energy (dashed lines)
and enthalpy (solid lines) when the transverse field is changed.
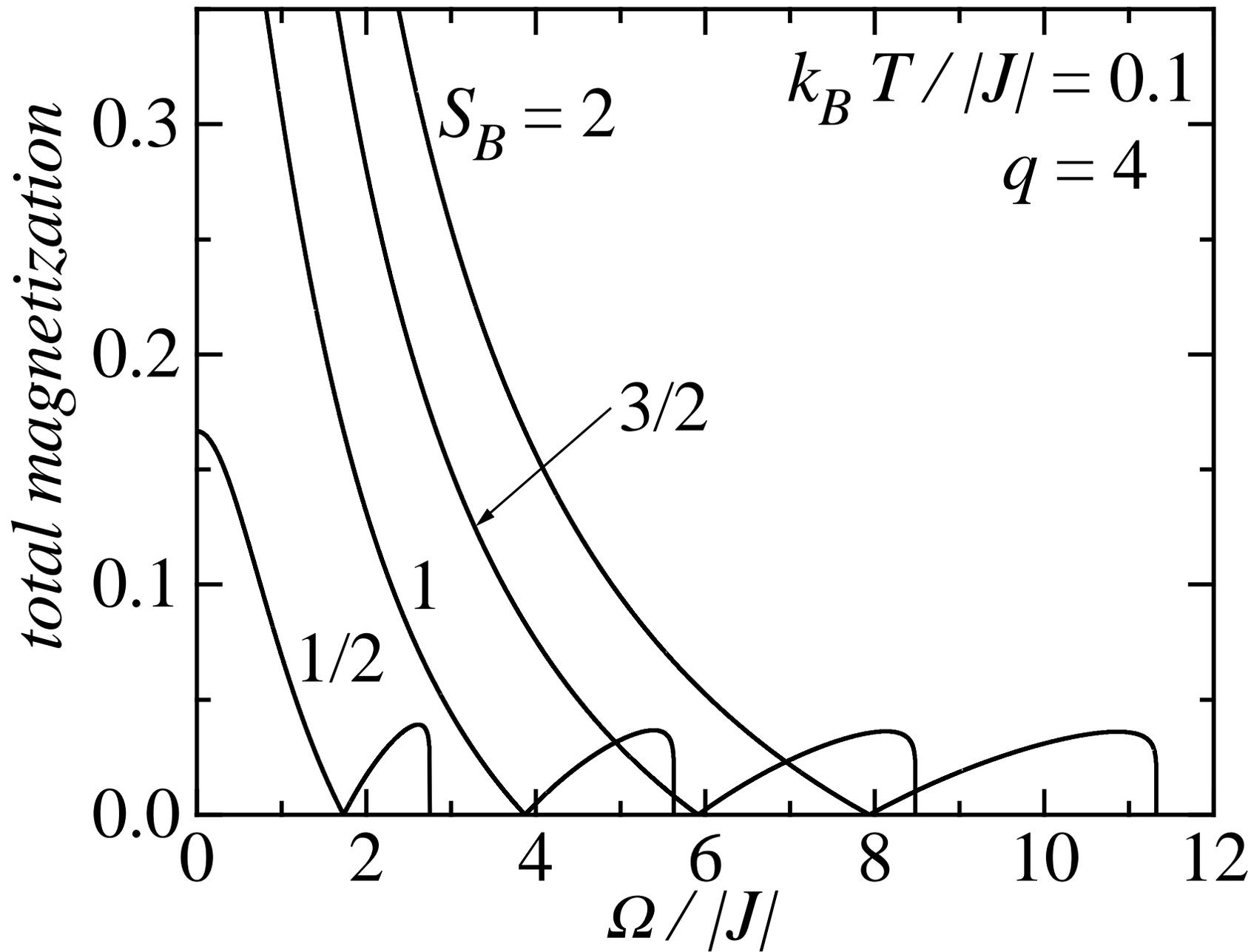
\item [Fig.5]
Total longitudinal magnetization per one site versus transverse field
for  $k_B T / |J| = 0.1$ and different spins of decorating atoms.
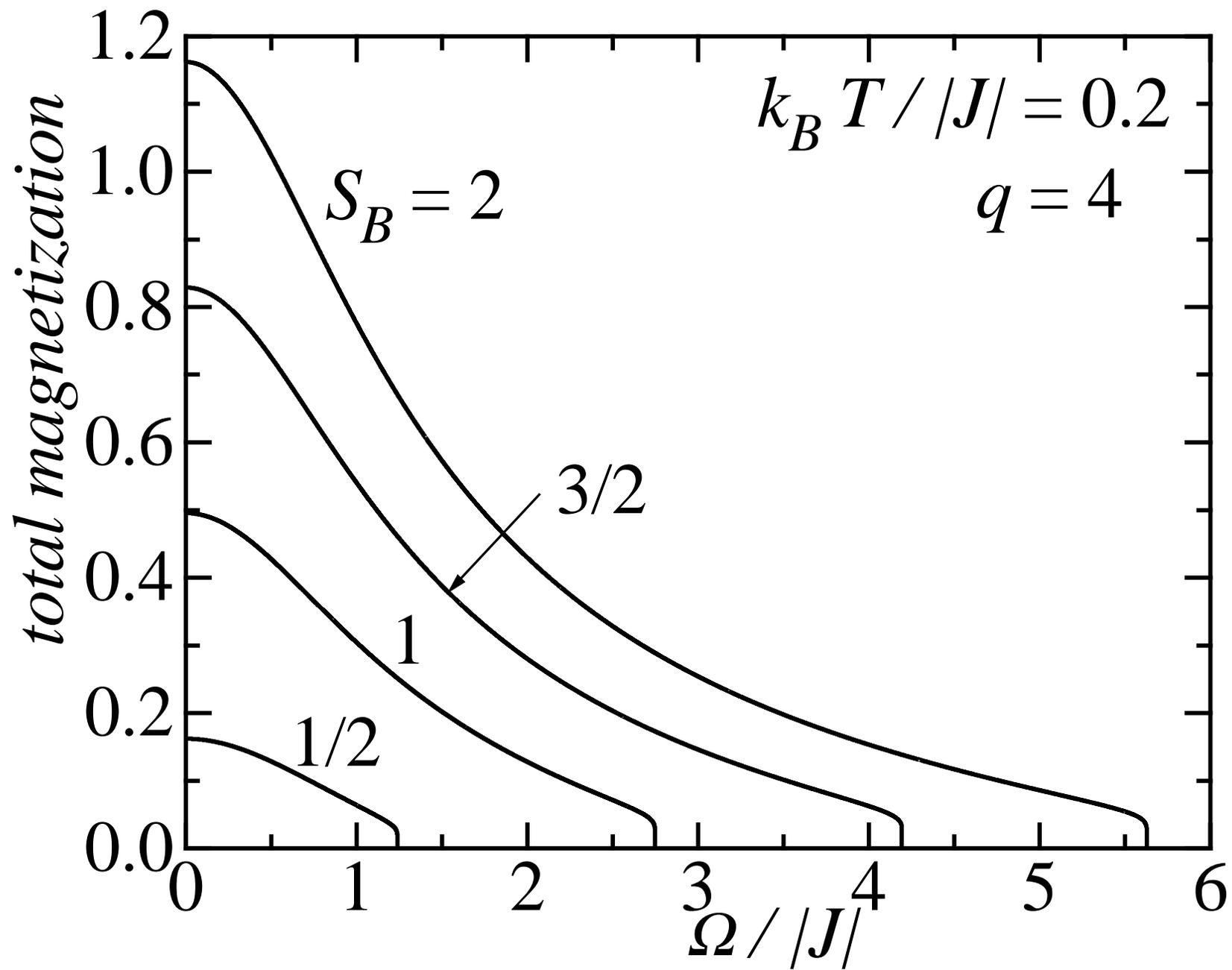
\item [Fig.6]
Total longitudinal magnetization versus transverse field
for  $k_B T / |J| = 0.2$ and different
spins of decorating atoms.
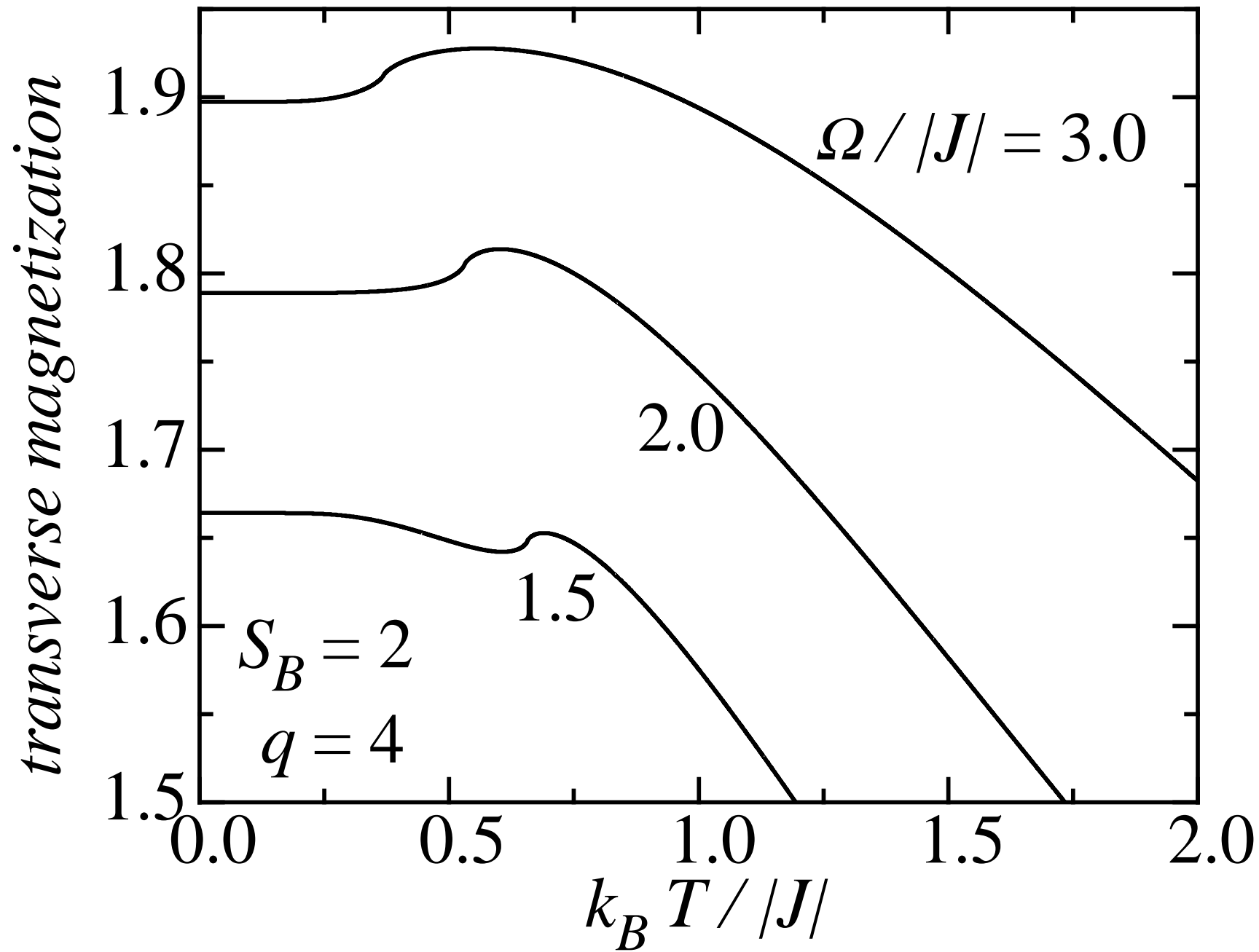
\item [Fig.7]
Transverse magnetization as a function of temperature for
$S_B = 2$ and different transverse fields.
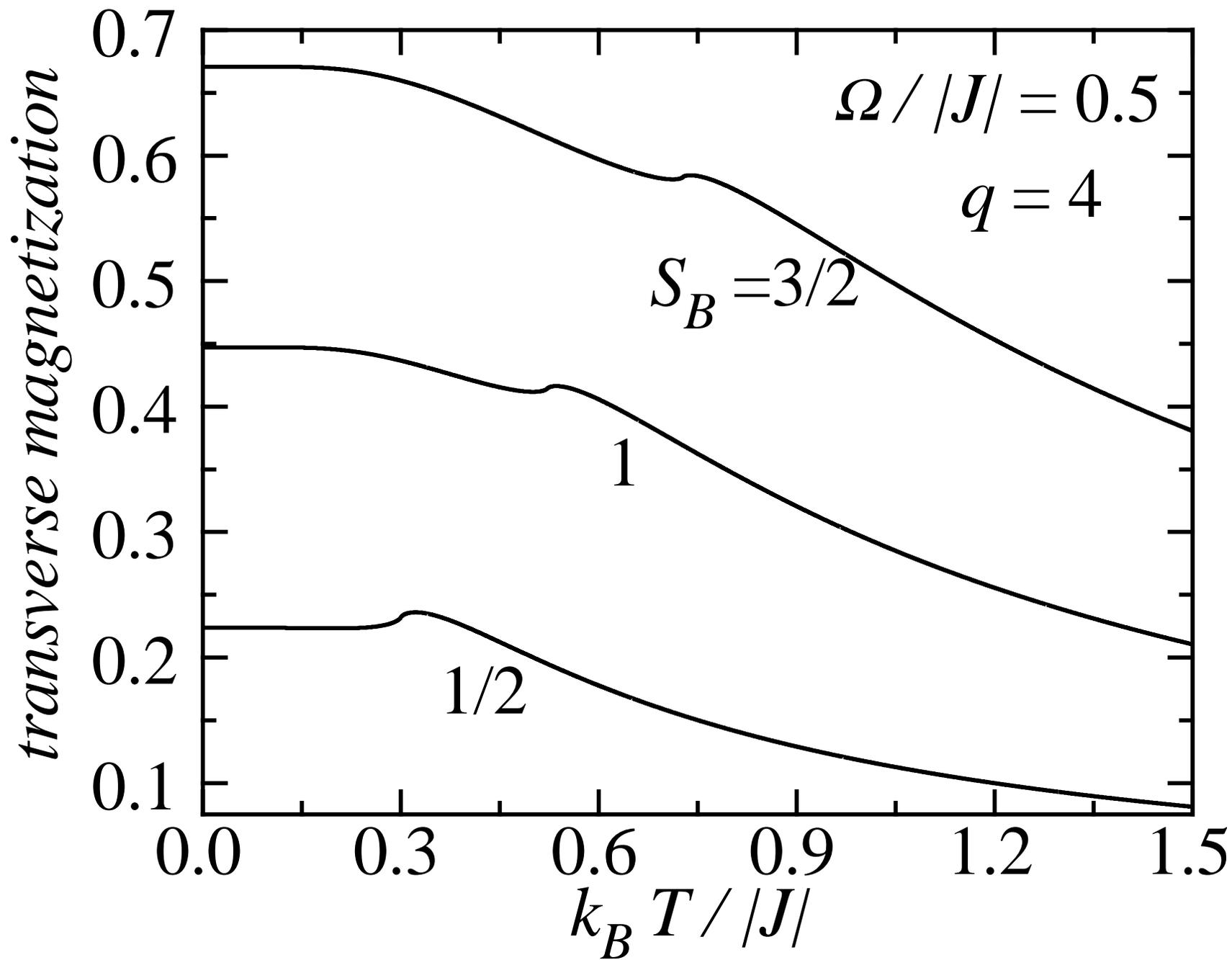
\item [Fig.8]
Thermal dependences of the transverse magnetization for several
spin values of decorating atoms, when the transverse field is
fixed ($\Omega/|J| = 0.5$).
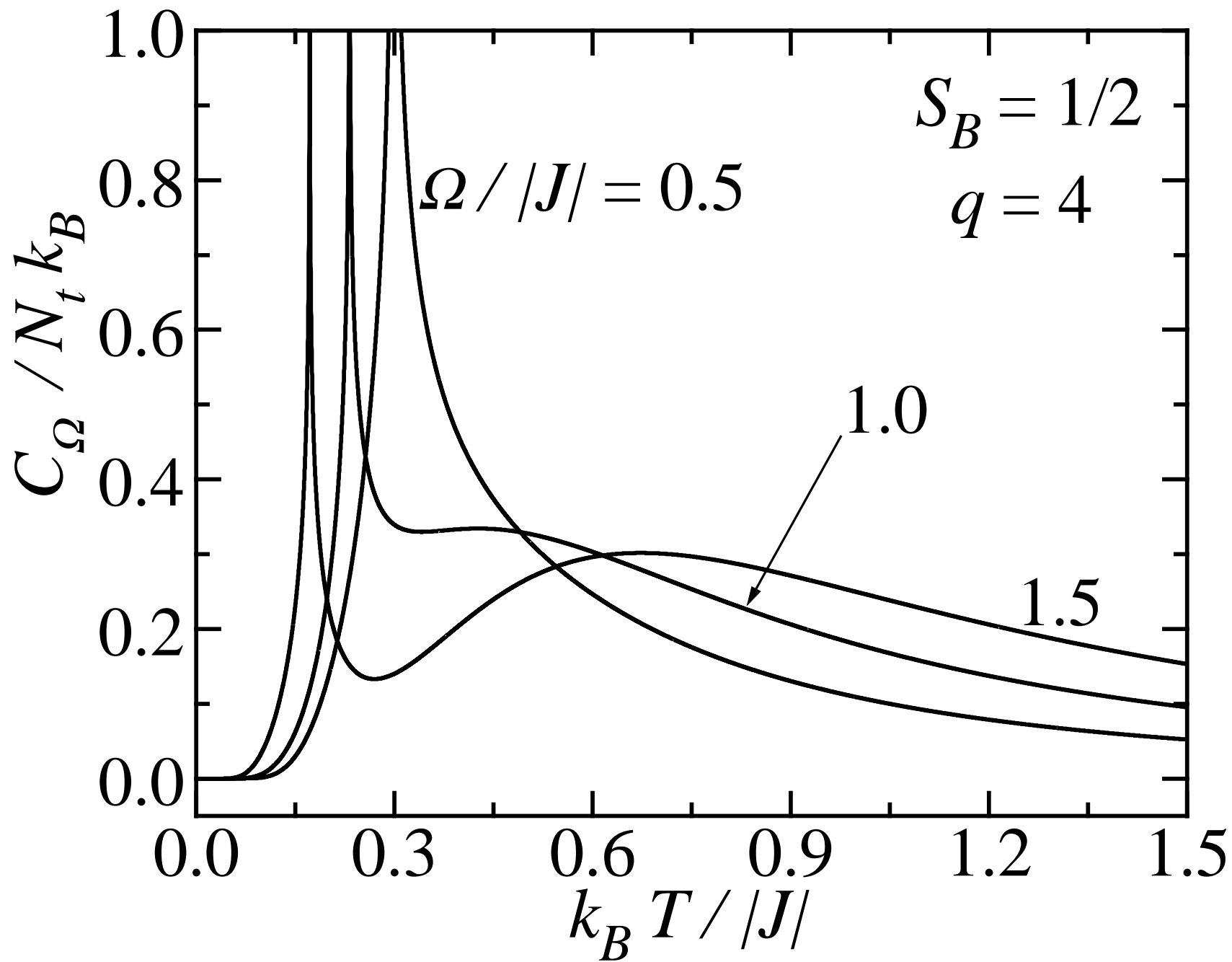
\item [Fig.9]
Specific-heat variations with the temperature for selected
values of the transverse field.
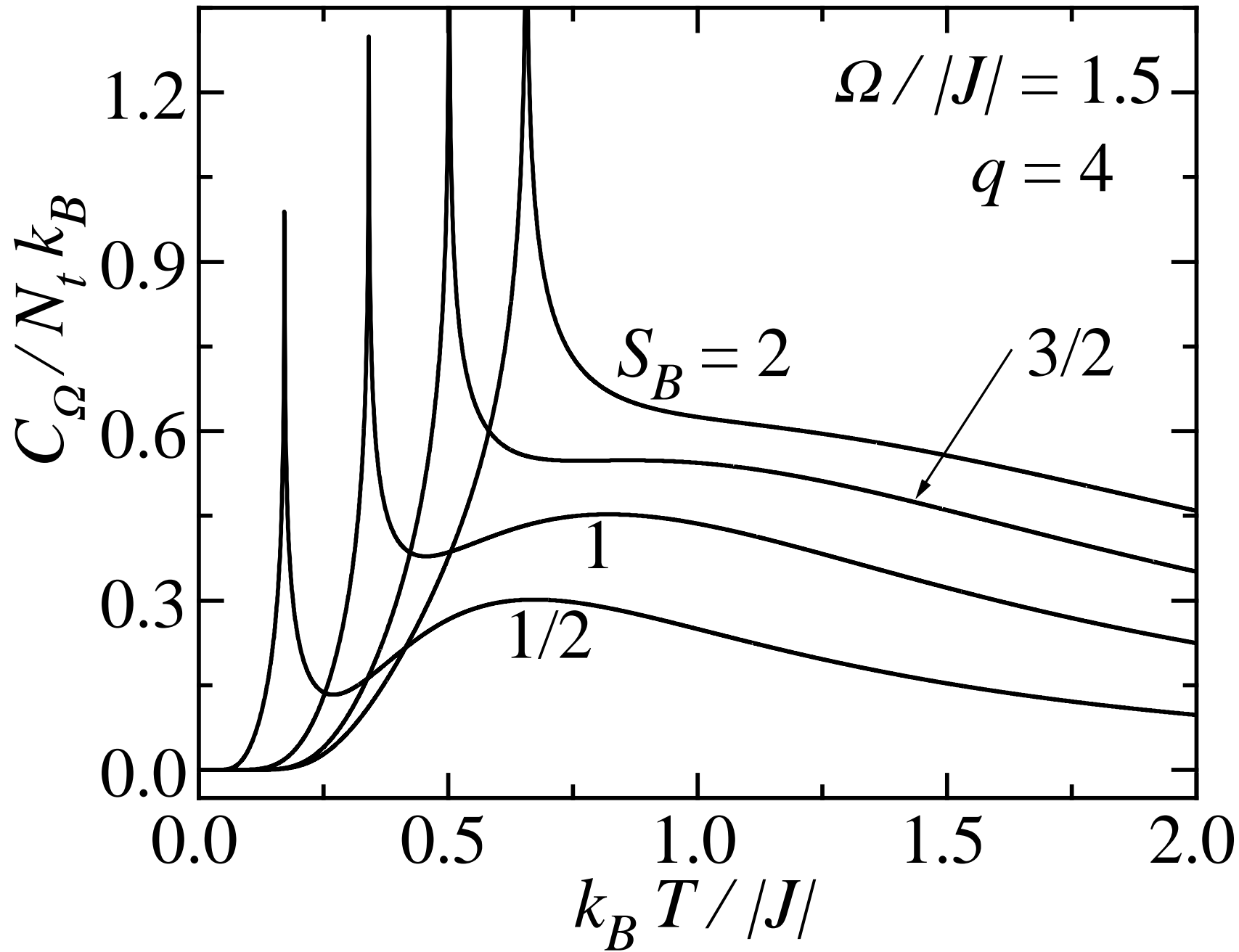
\item [Fig.10]
Thermal variations of the specific heat for different spin
cases of decorating atoms and the fixed transverse field
value ($\Omega / |J| = 1.5$).
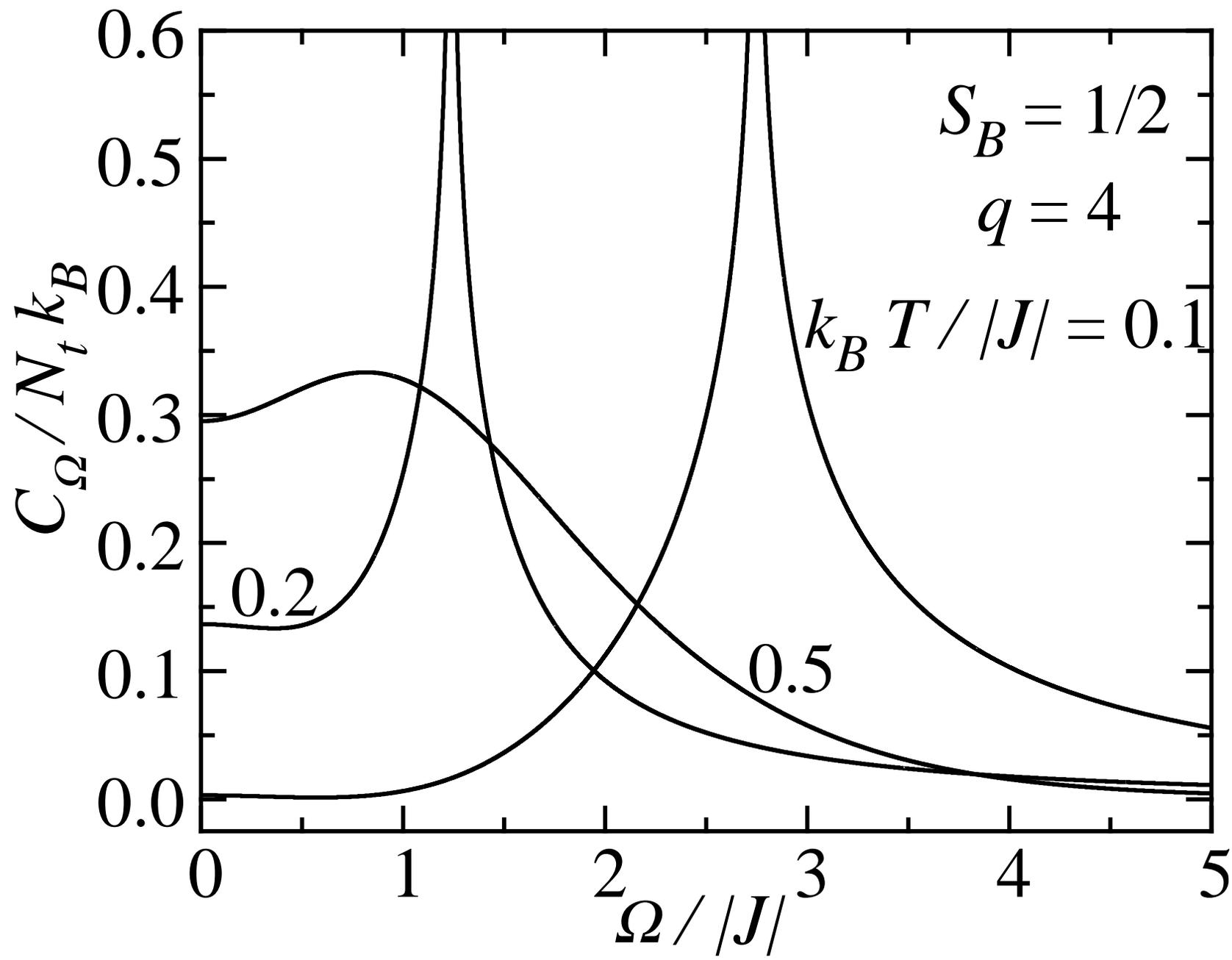
\item [Fig.11]
Specific heat dependence on the transverse field when the
temperature is changed.
\item [Fig.12]
Transverse susceptibility against the transverse field for
selected temperatures. In the insert, the comparison between
different spin cases is shown.
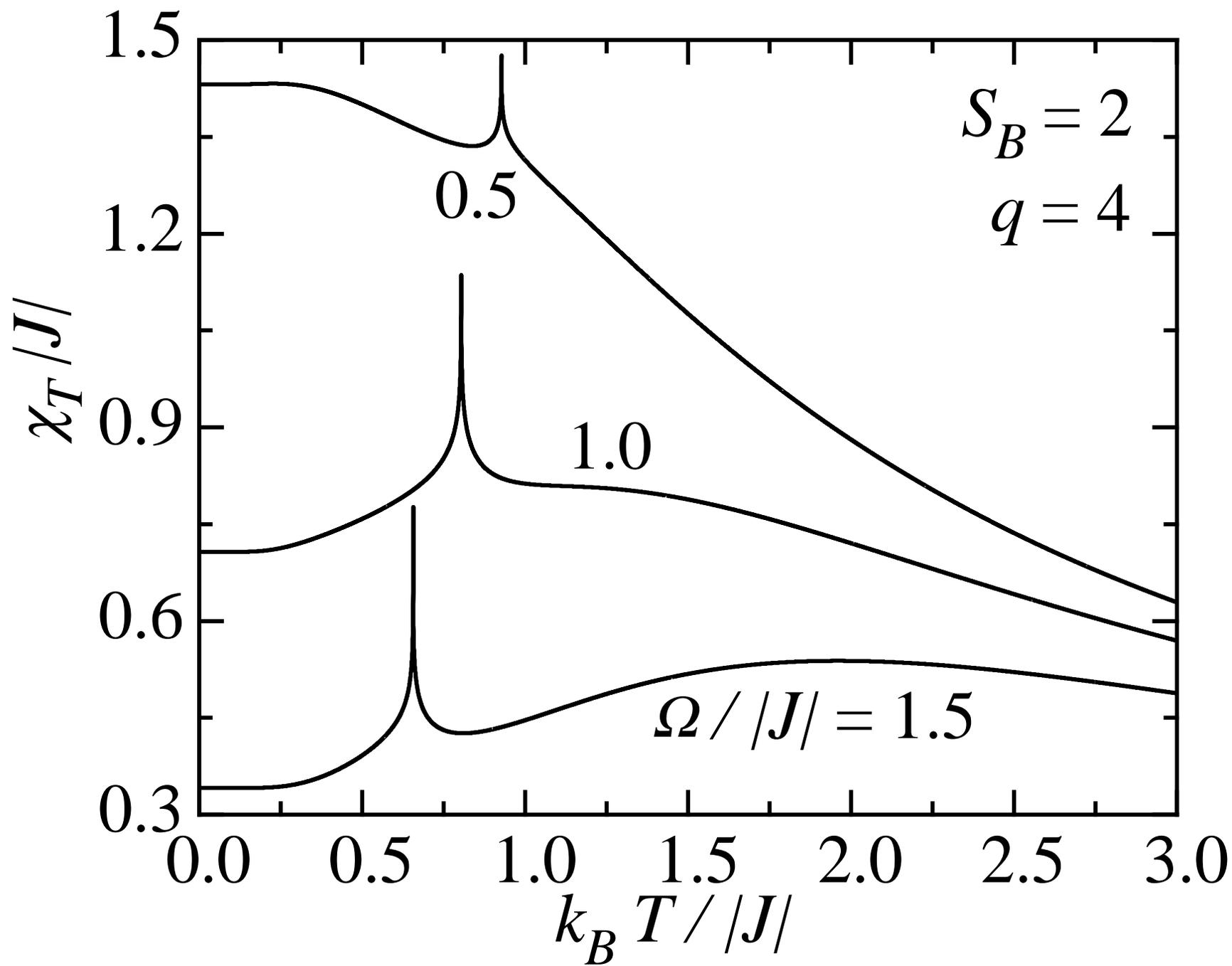
\item [Fig.13]
Transverse susceptibility dependence on the temperature
for the spin $S_B = 2$, when the transverse field is
changed.
\end{itemize}
\end{document}